\documentclass[reprint,amssymb,superscriptaddress,nofootinbib]{revtex4-2}
\usepackage[pdftex]{graphicx}
\usepackage{subfig}
\usepackage{amsmath,amsfonts,amssymb,mathtools}
\usepackage{tasks}
\usepackage{diagbox}
\usepackage{float}

\usepackage{float}
\usepackage{bbm,bm}
\usepackage[normalem]{ulem}
\usepackage{comment}
\usepackage{physics}
\usepackage{xcolor}
\usepackage[utf8]{inputenc}
\usepackage{csquotes}
\usepackage{pgfplots}
\pgfplotsset{compat=1.18}
\definecolor{blueprl}{RGB}{46,48,146}
\usepgfplotslibrary{groupplots}

\usepackage{lipsum}

\def\ANU{Centre for Quantum Computation and Communication Technology, Department of Quantum Science, Australian National University, Canberra, ACT 2601, Australia.}

\def\Astarnew{A*STAR Quantum Innovation Centre (Q.InC), Institute of Materials Research and Engineering (IMRE), Agency for Science Technology and Research (A*STAR), 2 Fusionopolis Way, 08-03 Innovis 138634, Singapore}

\begin{document}
%\title{Quantum information processing tasks with continuous variable photonic states}
\title{Verifying the security of a continuous variable quantum communication protocol via quantum metrology}
\author{Lorc{\'a}n O. Conlon$^{\dagger}$}
\email{lorcanconlon@gmail.com}
%\author{LC}
\affiliation{\ANU}
\affiliation{\Astarnew}
\author{Biveen Shajilal$^{\dagger}$}
\affiliation{\ANU}
\affiliation{\Astarnew}
\author{Angus Walsh}
\affiliation{\ANU}
\author{Jie Zhao}
\affiliation{\ANU}
\author{Jiri Janousek}
\affiliation{\ANU}
\author{Ping Koy Lam}
\affiliation{\ANU}
\affiliation{\Astarnew}
\author{Syed M. Assad}
\affiliation{\ANU}
\affiliation{\Astarnew}
%\affiliation{\NTU}
%\date{\today}

\begin{abstract}
Quantum mechanics offers the possibility of unconditionally secure communication between multiple remote parties. Security proofs for such protocols typically rely on bounding the capacity of the quantum channel in use. In a similar manner, Cram{\'{e}}r-Rao bounds in quantum metrology place limits on how much information can be extracted from a given quantum state about some unknown parameters of interest. In this work we establish a connection between these two areas. We first demonstrate a three-party sensing protocol, where the attainable precision is dependent on how many parties work together. This protocol is then mapped to a secure access protocol, where only by working together can the parties gain access to some high security asset. Finally, we map the same task to a communication protocol where we demonstrate that a higher mutual information can be achieved when the parties work collaboratively compared to any party working in isolation.
\end{abstract}

\maketitle
\def\thefootnote{$^\dagger$}\footnotetext{These authors contributed equally to this work}\def\thefootnote{\arabic{footnote}}

\begin{figure*}[t]
\includegraphics[width=0.9\textwidth]{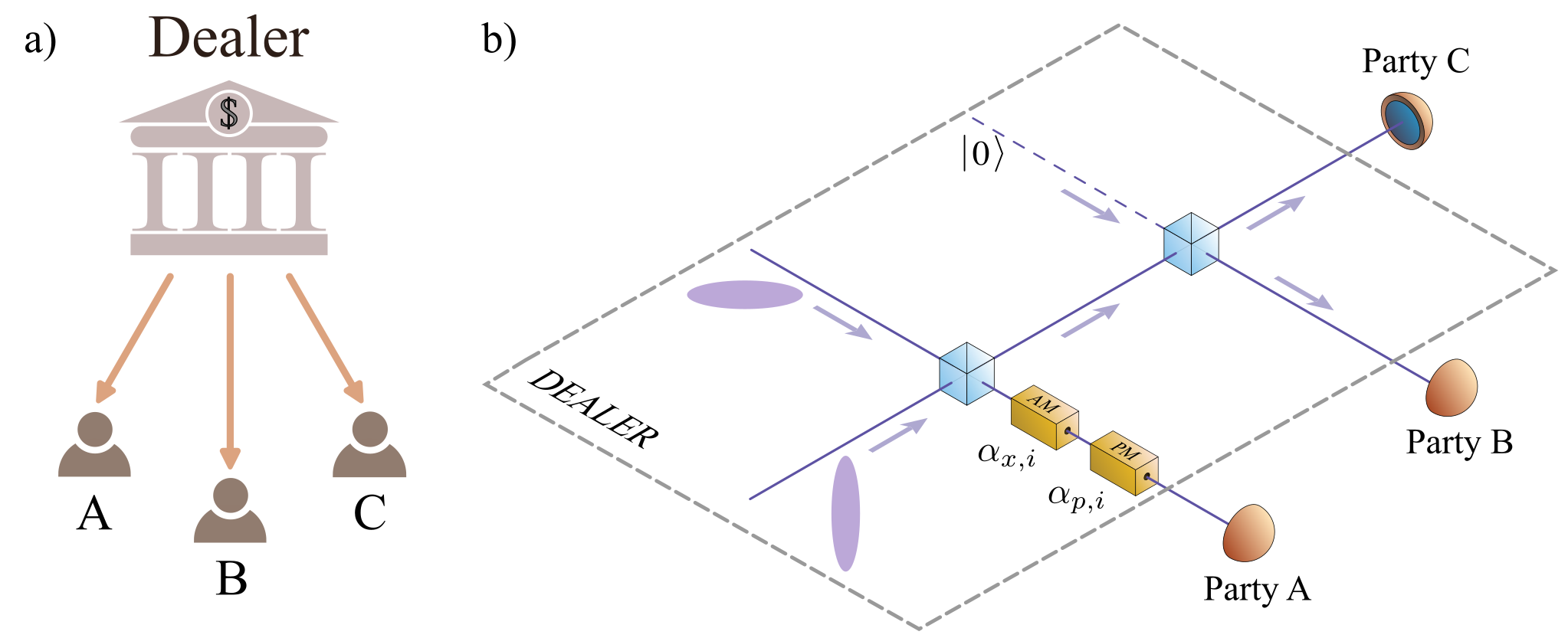}
\caption{\textbf{Schematic of the experimental set-up for demonstrating metrological tasks with a connection to secure communication.} a) Conceptual schematic of the experiment. Highly sensitive information, depicted here as access to a bank account, is shared among three parties in such a manner that only trusted groups of parties can access the information. b) Experimental set-up. The dealer (e.g. the banker in a)) mixes two squeezed states on a 50:50 beamsplitter to generate a two mode squeezed vacuum (TMSV) state. One mode is split on a second 50:50 beamsplitter and these modes are sent to parties B and C. The dealer implements a displacement $(\alpha_{x,i},\alpha_{p,i})$ in the $\hat{x}$ and $\hat{p}$ quadratures on the final arm of their state before this mode is sent to party A. Depending on whether the parties work cooperatively or independently differing amounts of information can be extracted about the unknown displacements $(\alpha_{x,i},\alpha_{p,i})$. AM and PM represent amplitude and phase modulators respectively.}
\label{fig:schematic1}
\end{figure*}
\section{Introduction}

Non-classical correlations have been shown to enable a range of quantum-enhanced tasks~\cite{brukner2004bell,masanes2006all}. One such example is quantum metrology, which utilises quantum resources to achieve a better measurement sensitivity than is classically possible. This could be in the form of quantum probe states~\cite{leibfried2004toward,kacprowicz2010experimental,daryanoosh2018experimental,pedrozo2020entanglement,marciniak2021optimal,guo2020distributed}, or quantum-enhanced measurements~\cite{roccia2017entangling,hou2018deterministic,conlon2023approaching,conlon2023discriminating}. Indeed, the connection between entanglement and quantum enhanced sensing has been widely studied~\cite{pezze2009entanglement,hyllus2010not,krischek2011useful,strobel2014fisher,toth2014quantum,toth2018quantum}. More recently, it has been shown that quantum metrology tasks can be used to witness steering~\cite{yadin2021metrological}. The connection between Bell correlations~\cite{bell1964einstein} and quantum sensing has also been investigated~\cite{frowis2019does,niezgoda2021many}.

Quantum resources also offer the promise of unconditionally secure communication. This can be done through quantum key distribution (QKD), which involves the sharing of a secret key between two parties in a manner such that no possible malicious evesdropper can access the key. The first such protocol was introduced by Bennett and Brassard in 1984, where information was encoded in the polarisation degree of freedom of photons~\cite{bennett2020quantum}. An entanglement-based QKD protocol was later proposed by Artur Ekert in 1991~\cite{ekert1991quantum}. Over the subsequent decades, there has been much progress towards QKD networks in various scenarios~\cite{vallone2015experimental, liao2017satellite, liao2018satellite, stucki2011long,sasaki2011field,dynes2019cambridge,yin2020entanglement,erkilicc2023surpassing}. A related area, which also uses the laws of quantum mechanics for its security, is that of secret sharing. This involves the sharing of either classical or quantum information between a number of trusted parties. There has been a great deal of work on the secret sharing of both classical and quantum information using both discrete variable (DV)~\cite{karlsson1999quantum,cleve1999share,tittel2001experimental,xiao2004efficient,zhang2005multiparty} and continuous variable (CV) systems~\cite{tyc2002share,lance2003continuous,lance2004tripartite,lance2005continuous,kogias2017unconditional,zhou2018quantum,grice2019quantum,wu2020passive,liao2021quantum}. Secret sharing can be used to ensure that only trusted parties gain access to highly confidential information such as bank accounts or missile launch codes.
 
In this work we design and implement a variant of conventional secret sharing protocols, allowing only certain sets of trusted parties to access confidential information. We use techniques from quantum metrology to bound the information which can be attained by the untrusted parties. Our protocol involves distributing a quantum state with unknown displacements among three parties. When considering quantum sensing, we will treat the unknown displacements as parameters to be estimated and when considering quantum communication, we will treat the unknown displacements as the secret information encoded in the quantum state. This enables us to draw a direct connection between these two tasks. %Although entanglement~\cite{takei2005high,chrzanowski2014measurement}, steering~\cite{ou1992realization,handchen2012observation} and Bell violations~\cite{thearle2018violation} have been observed many times over many years for CV systems, the connection of these non-classical correlations to other areas of quantum information, namely metrology and secure communication, remains important from a practical viewpoint. %In order to implement these quantum information processing tasks we require additional displacements on one mode of the two-mode state, introducing additional experimental complexity compared to traditional experiments witnessing quantum correlations.

The layout of this paper is as follows. In section~\ref{sec:prelim} we introduce the preliminary material needed. In sections~\ref{sec:theoryres} and \ref{sec:SS} we present our theoretical and experimental results respectively. Finally, we conclude the paper in section~\ref{sec:conc}
\section{Results}
\subsection{Preliminary material}
\label{sec:prelim}
A CV quantum state can be described by observables in an infinite dimensional Hilbert space which have a continuous spectrum of possible eigenvalues~\cite{weedbrook2012gaussian}. Gaussian states are those with Gaussian measurement statistics.
In our experiments, the variables of interest are the amplitude and phase quadratures of the electromagnetic field, denoted $\hat{x}$ and $\hat{p}$ respectively. These quadrature variables are defined in terms of the creation and annihilation operators, denoted $\hat{a}$ and $\hat{a}^\dagger$ respectively, as $\hat{x}=\hat{a}+\hat{a}^\dagger$ and $\hat{p}=-\mathrm{i}(\hat{a}-\hat{a}^\dagger)$. As the creation and annihilation operators satisfy $[\hat{a},\hat{a}^\dagger]=1$, the $\hat{x}$ and $\hat{p}$ quadrature operators satisfy $[\hat{x},\hat{p}]=2\mathrm{i}$. From the uncertainty principle~\cite{robertson1929uncertainty,heisenberg1985anschaulichen,arthurs1965simultaneous,arthurs1988quantum}, the variances in both quadratures must satisfy
\begin{equation}
\label{eq:ucprincip}
\Delta\hat{x}^2\Delta\hat{p}^2\geq1\;,
\end{equation}
where $\Delta\hat{x}^2$ is the variance in measuring the $\hat{x}$ quadrature and similarly for $\Delta\hat{p}^2$.

%. 
There do exist non-classical Gaussian states which can violate this uncertainty principle in certain settings. For example, by mixing two squeezed vacuum states on a 50:50 beam splitter one can create a two mode squeezed vacuum (TMSV) state. The two modes of this state are correlated in the $\hat{x}$ quadrature and anti-correlated in the $\hat{p}$ quadrature. With sufficiently high squeezing, low antisqueezing, and low loss, this state can be entangled or steerable in the $\hat{x}$ and $\hat{p}$ quadratures. With such states, it is possible to achieve a measurement sensitivity better than what is allowed by Eq.~\eqref{eq:ucprincip}, see e.g Ref.~\cite{bradshaw2018ultimate}. The non-classical correlations of the TMSV enable uniquely quantum mechanical tasks such as quantum teleportation~\cite{furusawa1998unconditional,zhao2023enhancing}, quantum illumination~\cite{tan2008quantum,bradshaw2021optimal}, quantum enhanced sensing~\cite{bradshaw2018ultimate,bradshaw2017tight,assad2020accessible,steinlechner2013quantum,d2001using} and QKD~\cite{ralph1999continuous,ralph2000security} to be carried out.

\subsection{Theoretical Results}
\label{sec:theoryres}

We consider the set-up shown in Fig.~\ref{fig:schematic1}. The classical information that the dealer wishes to share with the three parties are displacements in the $\hat{x}$ and $\hat{p}$ quadratures, denoted as $\alpha_{x,i}$ and $\alpha_{p,i}$ respectively\footnote{The rational behind using subscript $i$ will become evident later in the manuscript}. In an ideal experimental implementation of our protocol, the dealer prepares a TMSV state and introduces displacements on one arm. After mixing one mode of the TMSV on a second 50:50 beamsplitter with an ancilla vacuum state, the three parties in Fig.~\ref{fig:schematic1} share a continuous variable state with the mean vector $\begin{pmatrix}0&0&0&0&\alpha_{x,i}&\alpha_{p,i}\end{pmatrix}^T\;$ where the modes are ordered as $\begin{pmatrix}\langle \hat{x}_C\rangle&\langle \hat{p}_C\rangle&\langle \hat{x}_B\rangle&\langle \hat{p}_B\rangle&\langle \hat{x}_A\rangle&\langle \hat{p}_A\rangle\end{pmatrix}^T\;$, where $\langle \hat{i}_j\rangle$ denotes the expectation value of the $\hat{i}$ quadrature for party $j$.
%\begin{equation}
%\begin{pmatrix}
%0\\
%0\\
%0\\
%0\\
%\alpha_x\\
%\alpha_p
%\end{pmatrix}\;,
%\end{equation}
The corresponding covariance matrix is
\begin{widetext}
\begin{equation}
\label{eqthreemsv}
\begin{pmatrix}
\text{cosh}(r)^2&0&-\text{sinh}(r)^2&0&-\sqrt{2}\text{cosh}(r)\text{sinh}(r)&0\\
0&\text{cosh}(r)^2&0&-\text{sinh}(r)^2&0&\sqrt{2}\text{cosh}(r)\text{sinh}(r)\\
-\text{sinh}(r)^2&0&\text{cosh}(r)^2&0&\sqrt{2}\text{cosh}(r)\text{sinh}(r)&0\\
0&-\text{sinh}(r)^2&0&\text{cosh}(r)^2&0&-\sqrt{2}\text{cosh}(r)\text{sinh}(r)\\
-\sqrt{2}\text{cosh}(r)\text{sinh}(r)&0&\sqrt{2}\text{cosh}(r)\text{sinh}(r)&0&\text{cosh}(2r)&0\\
0&\sqrt{2}\text{cosh}(r)\text{sinh}(r)&0&-\sqrt{2}\text{cosh}(r)\text{sinh}(r)&0&\text{cosh}(2r)
\end{pmatrix}\;,
\end{equation} 
\end{widetext}
where $r$ is the squeezing parameter. We now wish to investigate various quantum information tasks which can be achieved with this state and the connection between these tasks.

\subsubsection{Multiparameter estimation -- Simultaneous estimation of displacements in both quadratures}
We first compute the theoretical limits on how precisely the displacements shown in Fig.~\ref{fig:schematic1} can be simultaneously estimated. This is done for one party working alone, parties working in pairs, and all three parties working together. For a single party working alone, we shall use the Holevo Cram\'{e}r-Rao bound (HCRB) to evaluate the precision which can be achieved. 
This is done as the HCRB provides the ultimate limit on the variance which can be achieved in multiparameter estimation~\cite{holevo1973statistical,holevo2011probabilistic}. In general, the HCRB may require an entangling collective measurement on infinitely many copies of the probe state to be saturated~\cite{kahn2009local, yamagata2013quantum,yang2019attaining,conlon2022gap}, suggesting that other Cram\'{e}r-Rao bounds may be more experimentally relevant~\cite{nagaoka2005new,nagaoka2005generalization,conlon2021efficient}. However, in the specific case of estimating Gaussian displacements, the HCRB can be saturated by linear measurements~\cite{holevo2011probabilistic}. When considering two and three parties working together, we shall evaluate the precision attainable by a specific measurement strategy. This is done to allow us to compare the experimentally attained two and three party precisions to the ultimate theoretical limits on the precision attainable by a single party.

The average mean squared error (MSE) that the party or parties can achieve when estimating $\alpha_{x,i}$ is given by
\begin{equation}
\text{MSE}_x=(\alpha_{x,i}-\tilde{\alpha}_{x,i})^2\;,
\end{equation}
where we use a tilde to denote the estimated value and $\text{MSE}_p$ is defined similarly. When $m$ parties work together, we will denote the average MSE with which $\alpha_{x,i}$ and $\alpha_{p,i}$ can be measured as $v_{\alpha_x,m}$ and $v_{\alpha_p,m}$ respectively. Clearly we have $v_{\alpha_x,1}\geq v_{\alpha_x,2}\geq v_{\alpha_x,3}$. 

\noindent
\textit{One party MSE} \newline\noindent
 From Fig.~\ref{fig:schematic1}, it is evident that only party A will be able to access any information about the unknown displacements when working in isolation. Without any information from the other two parties, party A will receive a displaced thermal state, obtained by tracing out the first and second modes of the shared state in Eq.~\eqref{eqthreemsv}. In the ideal case, party A obtains a thermal state with variance $\text{cosh}(2r)$ in both quadratures. More generally, we may have a slight asymmetry between the two quadratures, and so we write the covariance matrix of the state accessible by party A as
\begin{equation}
\label{eq:genthermal}
\begin{pmatrix}
v_{1}&0\\
0&v_{2}
\end{pmatrix}=
\begin{pmatrix}
1+2n_{1}&0\\
0&1+2n_{2}
\end{pmatrix}\;,
\end{equation}
where $n_i$ characterises the thermal variance. If $n_1=n_2$, then this quantity represents the mean thermal photon number. Note that without loss of generality we can assume that $v_{1}\geq v_{2}$. In Appendix~\ref{apend:HCRB} we show that, for this state, the HCRB for estimating the amplitude and phase displacements simultaneously is 
\begin{equation}
\label{eqhcrbmain}
v_{\alpha_x,1}+v_{\alpha_p,1}\geq\mathcal{C}_\text{H}=4+2n_{1}+2n_{2}\;.
\end{equation}
Note that this MSE is normalised for the number of probe states used. In the scenario shown in Fig.~\ref{fig:schematic1}, this represents the smallest possible average sum of the MSE that party A can attain using an unbiased estimator.

\noindent
\textit{Two party MSE}\newline\noindent
When considering two parties working together, we will not use the HCRB to bound the MSE. Rather, we shall directly compute the MSE attainable when using homodyne detection as this is what was implemented in our experiment. Let us consider the ideal case, with no loss or thermal noise\footnote{When fitting to experimental data we take such imperfections into account}. In this scenario, parties A and B (also parties A and C) share the following covariance matrix
\begin{widetext}
\begin{equation}
\label{eqtmsv}
\begin{pmatrix}
\text{cosh}(r)^2&0&\sqrt{2}\text{cosh}(r)\text{sinh}(r)&0\\
0&\text{cosh}(r)^2&0&-\sqrt{2}\text{cosh}(r)\text{sinh}(r)\\
\sqrt{2}\text{cosh}(r)\text{sinh}(r)&0&\text{cosh}(2r)&0\\
0&-\sqrt{2}\text{cosh}(r)\text{sinh}(r)&0&\text{cosh}(2r)
\end{pmatrix}\;.
\end{equation} 
\end{widetext}
Interestingly, by recombining the measurement results of parties A and B (scaling the measurement results of party B by an optimised factor), it is possible to always achieve a MSE of $v_{\alpha_x,2}=v_{\alpha_p,2}=1$, regardless of the squeezing level. In this case the estimator used for the unknown Gaussian displacement is
\begin{equation}
\tilde{\alpha}_{x}=x_A-g_Bx_B\;,
\end{equation}
where we use $x_j$ to denote the $\hat{x}$ quadrature measurement results for party $j$ and $g_B$ is a constant chosen to minimise the MSE. As $g_B$ depends on the experimental parameters, such as $r$, it must be optimised for every data point. Quantities for the $\hat{p}$ quadrature are similarly defined.

For a fair comparison with the single party case, in the two party case we measure each quadrature with half of the total states, which increases the MSE by a factor of 2, giving
\begin{equation}
\label{eqMSEtwoparty}
v_{\alpha_x,2}+v_{\alpha_p,2}\geq4\;.
\end{equation}
In any experimental implementation with imperfections, the variance achieved by any two parties can only be larger than this.

\begin{figure}[t!]
\includegraphics[width=0.5\textwidth]{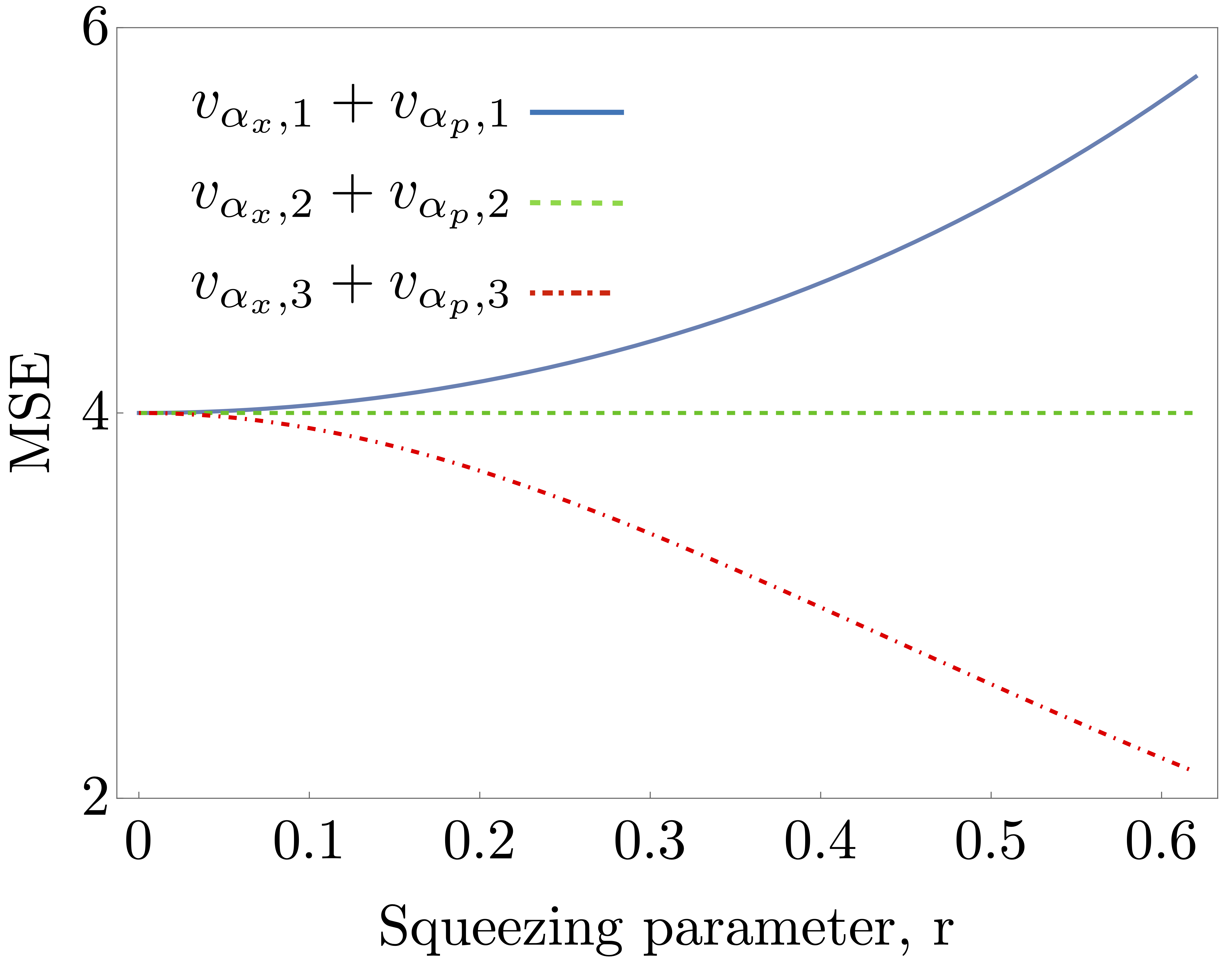}
\caption{\textbf{Theoretical limits on the MSE in an ideal experiment when we restrict to the set-up shown in Fig.~\ref{fig:schematic1}.} We show how the attainable MSE changes as a function of the squeezing level for one party working independently, two parties working together and three parties working together.}
\label{fig:th1}
\end{figure}

\noindent
\textit{Three party MSE}\newline\noindent
As before, we consider the ideal case, with no loss or thermal noise. In this scenario, if all three parties measure the same quadrature, parties B and C can recombine their results, scaled by a factor $\sqrt{2}$, so that all three parties effectively share an ideal two mode squeezed vacuum state. Hence in this case, the estimator that is used is
\begin{equation}
\tilde{\alpha}_{x}=x_A-g_{BC}(x_B-x_C)/\sqrt{2}\;,
\end{equation}
where $g_{BC}$ is a constant optimised to minimise the MSE, and a similar estimator is used for $\alpha_{p}$. Note that when there are no experimental imperfections, the optimal value of $g_{BC}$ is $g_{BC}=\text{tanh}(2r)$~\cite{reid2009colloquium}. However, with experimental imperfections $g_{BC}$ must be optimised for every data point. Using this information, it is easy to calculate that in the ideal case all three parties can achieve a MSE of
\begin{equation}
\label{eqthreepartyMSE}
v_{\alpha_x,3}+v_{\alpha_p,3}\geq\frac{8}{\text{e}^{2r}+\text{e}^{-2r}}\;,
\end{equation}
whic goes to 0 in the limit of infinite squeezing. A comparison of the MSE which can be attained by the different combinations of parties is shown in Fig.~\ref{fig:th1}.

\subsubsection{Secure access protocol}
\label{sec:SSprotocol}
We are now in a position to introduce our secure access protocol. For this we draw a comparison with existing quantum secret sharing protocols. Note that we are considering the sharing of classical information using quantum states, as opposed to sharing the quantum state itself. In conventional $(k,n)$ secret sharing, the dealer encodes information in such a way that any $k$ parties, out of the total $n$ parties, can work together to reconstruct the information encoded by the dealer. The remaining $n-k$ parties cannot access any information. In practice, however, such perfect secret sharing is hampered by experimental imperfections and the finite entanglement available. We consider the protocol secure, only if the groups of $k$ parties acquire more information than the groups of $n-k$ parties. \footnote{Note that we can also consider approximate secret sharing, where $k$ parties acquire some approximate information about the secret, as do the remaining $n-k$ parties~\cite{crepeau2005approximate,PhysRevA.108.012425}.}

Our protocol differs slightly from existing protocols. In the $i$th round of the protocol the dealer implements displacements $\bm{\alpha}_i=(\alpha_{x,i},\alpha_{p,i})$. In each round the parties, either working together or independently, estimate either $\alpha_{x,i}$ or $\alpha_{p,i}$. As different measurements are needed to acquire information about $\alpha_{x,i}$ and $\alpha_{p,i}$, the parties must announce at the end of the entire protocol which parameter they were trying to measure in each round, $\alpha_{x,i}$ or $\alpha_{p,i}$. Any rounds in which the parties measured different parameters can then be discarded. We use $(\alpha_{x},\alpha_{p})$ to denote the two-dimensional vector of $\alpha_{x,i}$ and $\alpha_{p,i}$ values that are not discarded. After the reconciliation step, if we have $M_x$ remaining rounds of data in which $\alpha_{x,i}$ was measured by all parties, the average MSE that the parties achieve is given by
\begin{equation}
\text{MSE}_x=\frac{1}{M_x}\sum_{i}(\alpha_{x,i}-\tilde{\alpha}_{x,i})^2\;,
\end{equation}
 and $\text{MSE}_p$ is defined similarly. 

As discussed in the previous section, given a certain input state, the dealer can use bounds from quantum metrology to place limits on how small $v_{\alpha_x,m}$ and $v_{\alpha_p,m}$ can be. This allows the dealer to define a threshold average MSE $v_\text{T}$, below which the protocol is declared secure. In our setting, if we wish to ensure that no single party can access the trusted information, we shall refer to a protocol as $\delta$ secure if
\begin{equation}
\label{eq:security1}
\text{Pr}(v_{\alpha_x,1}+v_{\alpha_p,1}\leq v_\text{T})\leq\delta\;,
\end{equation} 
where $\text{Pr}(X)$ is the probability of the event $X$ occurring. Intuitively, $\delta$ represents the likelihood of any single party obtaining an average MSE below a certain threshold, or equivalently acquiring an amount of information about the displacements above a certain threshold. Although the MSE attainable by any party working independently (Eq.~\eqref{eqhcrbmain}) is larger than when the parties work together (Eqs.~\eqref{eqMSEtwoparty}, \eqref{eqthreepartyMSE}), this is only true statistically, i.e. on any given experimental run there is some finite probability that a party working alone predicts a value for $\alpha$ which is very close to the true value. Hence, in any practical setting with finite statistics, security can only be guaranteed up to some probability, Eq.~\eqref{eq:security1}. The MSE values attained follow a scaled $\chi^2$ distribution. When using $N$ probe states to measure each quadrature, if the mean MSE is denoted $\mu_\text{M}$, then the probability density of MSE's which will be attained is given by
\begin{equation}
\label{eq:pdf}
P(\text{MSE}=x)=\frac{2N}{\mu_\text{M}2^{N}\Gamma(N)}\bigg(\frac{2xN}{\mu_\text{M}}\bigg)^{N-1}\text{e}^{-\frac{xN}{\mu_\text{M}}}\;,
\end{equation}
see Appendix~\ref{apen:pdfderiv} for the derivation. 

Given the quantum state generated by the dealer, we can use the above equation to compute \mbox{$\text{Pr}(v_{\alpha_x,1}+v_{\alpha_p,1}\leq v_\text{T})$} for all possible $v_\text{T}$. From this definition alone, we can trivially choose $v_\text{T}=0$ to ensure that maximum security is achieved. However, in this case the protocol will never succeed, i.e. although any individual party cannot achieve an MSE of 0, with any practical resources, all of the parties working together also cannot achieve an MSE of 0. Thus, it is necessary to also define the success rate as
\begin{equation}
\label{eq:security2}
P_{s,2(3)}=\text{Pr}(v_{\alpha_x,2(3)}+v_{\alpha_p,2(3)}\leq v_\text{T})\;.
\end{equation} 
A good protocol should minimise $\delta$ and maximise $P_{s,2(3)}$.

\subsubsection{Attainable mutual information}
\label{sec:attainMI}
Finally, let us consider how to quantify, the amount of classical information that the dealer can share with these three parties. Assume the dealer chooses the displacements $\bm{\alpha}$ from a Gaussian distribution with variance $V_\text{dist}$. The parties involved in the protocol then attempt to estimate $\alpha_{x,i}$ and $\alpha_{p,i}$ as well as they possibly can. Averaging over all $(\alpha_{x},\alpha_{p})$ allows the correlation matrix, between the dealer and any number of parties to be constructed as
\begin{equation}
\label{eqcor}\sigma=
\begin{pmatrix}
V_\text{dist}&0&V_\text{dist}&0\\
0&V_\text{dist}&0&V_\text{dist}\\
V_\text{dist}&0&V_\text{dist}+v_{\alpha_x,m}&0\\
0&V_\text{dist}&0&V_\text{dist}+v_{\alpha_p,m}
\end{pmatrix}\;,
\end{equation} 
where $\sigma_{i,j}=\langle\tilde{\alpha}_i\tilde{\alpha}_j\rangle-\langle\tilde{\alpha}_i\rangle\langle\tilde{\alpha}_j\rangle$ and we use the mode ordering $(\tilde{\alpha}_{x,D},\tilde{\alpha}_{p,D},\tilde{\alpha}_{x,p},\tilde{\alpha}_{p,p})$ where (in an abuse of notation) the second subscript now denotes whether we are considering either the dealer or the players estimate. Thus, the first two diagonal elements represent the variance of the parameters to be estimated, and the second two diagonal elements represent this variance plus the error associated with the imperfect measurement of the unknown displacements. The off-diagonal elements represent the correlation between the dealer's estimate (i.e. the true value) and the estimate made by the players. Note that in this way multiple different covariance matrices can be constructed, between the dealer and any single party, between the dealer and any pair of parties and between the dealer and all three parties. From these covariance matrices the mutual information between the dealer and the different parties can be constructed. Let us make the assumption that $v_{\alpha_x,m}=v_{\alpha_p,m}=v_{\alpha,m}$. Then the mutual information can be calculated as
\begin{equation}
\label{eq:MI1}
\text{MI}=\text{log}(V_\text{dist}+v_{\alpha,m})-\text{log}(v_{\alpha,m})\;.
\end{equation}
Therefore, to achieve a mutual information of $c$ bits or more, we require a MSE less than or equal to
\begin{equation}
\label{eq:MI2}
v_{\alpha,m}=\frac{V_\text{dist}}{2^c-1}\;.
\end{equation}
From this equation and Eq.~\eqref{eq:pdf}, we can determine the probability of obtaining a mutual information above a certain value. 

This allows us to compare the mutual information when different numbers of parties work together. We note that, due to the asymmetry of our scheme, when parties C or B work individually or as a pair, they can access no information. Hence, we will ignore these combinations going forward. In this sense, we are not implementing \textquote{real} secret sharing, as not all subsets of two or more parties can access the secret information.

%The dealer can divide phase space into a total of $S$ squares, each of width $w$. The dealer then chooses a square at random, and communicates this information to the three parties by encoding the center coordinate of each square as $(\alpha_x,\alpha_p)$. Maybe the position of the squares can be announced only at a later stage / doesn't need to be announced at all ? 
%
%Does this raise an issue of the fact that we don't need to stay in the usual quantum metrology setting? I think so, one party working alone could have more information and so can do better than the HCRB.
%
%Maybe the dealer encodes information from a continuous probability distribution. We can look at the mutual information of the different squid-squads. 
%
%We can calculate the covariance matrix obtained in various cases and then use the mutual information.

\subsection{Experimental results}
\label{sec:SS}
In this section we will first describe our experimental set-up, and then verify that the quantum state we are using is entangled. We next present results for the sensing task described in the previous section. Finally, we connect these results to the secret sharing of classical information. Secret sharing using CV states has been investigated many times in the past~\cite{tyc2002share,lance2003continuous,lance2004tripartite,lance2005continuous,kogias2017unconditional,zhou2018quantum,grice2019quantum,wu2020passive,liao2021quantum} and so the idea is not novel in and of itself. However, we shall analyse the security through the HCRB. For DV systems, the analogy between secure quantum sensing and quantum secret sharing was noted in Ref.~\cite{huang2019cryptographic}. In this work security was obtained by preparing check states with a certain probability, as opposed to measuring conjugate parameters as is the case here.

\subsubsection{Experimental set-up}
A schematic of the experimental set-up is shown in Fig.~\ref{fig:schematic1} b). The dealer, generates two squeezed states and mixes them on a 50:50 beam splitter. Details on the squeezed light sources used can be found in Ref.~\cite{zhao2020high}. One mode of this state is subject to a second 50:50 beam splitter from which the two output modes are respectively distributed to parties B and C. A displacement in each quadrature is implemented on the remaining mode, before this is distributed to party A. In practice this displacement is implemented through an auxiliary beam which is amplitude and phase modulated and then mixed with party A's mode of the TMSV state on a 98:2 beam splitter. The aim of the three parties, either working together or independently, is to measure the displacements as accurately as possible. In our experiment, each party implements homodyne detection on either the $\hat{x}$ or $\hat{p}$ quadrature. However, in our security analysis, we do not place such restrictions on any party.

\subsubsection{Entanglement-enabled metrology task}
\label{ssEntWitness}

\begin{figure}[t]
\includegraphics[width=0.5\textwidth]{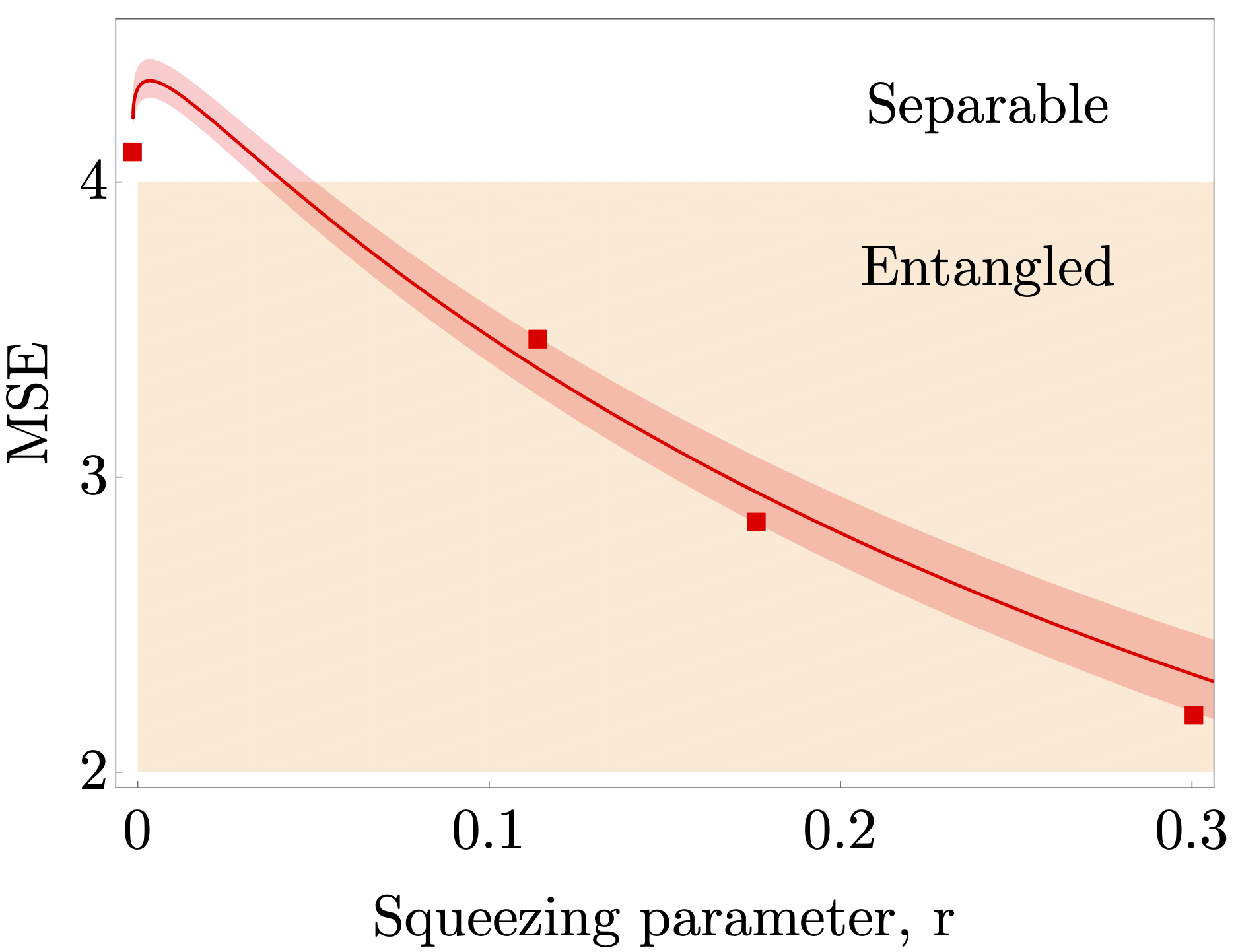}
\caption{\textbf{Entanglement-witnessing metrology task.} The experimental data corresponds to the MSE when estimating $(\alpha_x,\alpha_p)/\sqrt{2}$ using the estimator described in section~\ref{ssEntWitness}. The orange shaded region shows MSE values smaller than 4 which indicates that the parties share an entangled state. The $x-$axis shows the effective squeezing parameter. The alpha values being estimated on average have $\abs{\alpha}=0.2$. Each data point corresponds to at least $10^7$ measurement results and statistical error bars based on one standard deviations are smaller than the marker size.}
\label{fig:exp1}
\end{figure}

\begin{figure}[t]
\includegraphics[width=0.5\textwidth]{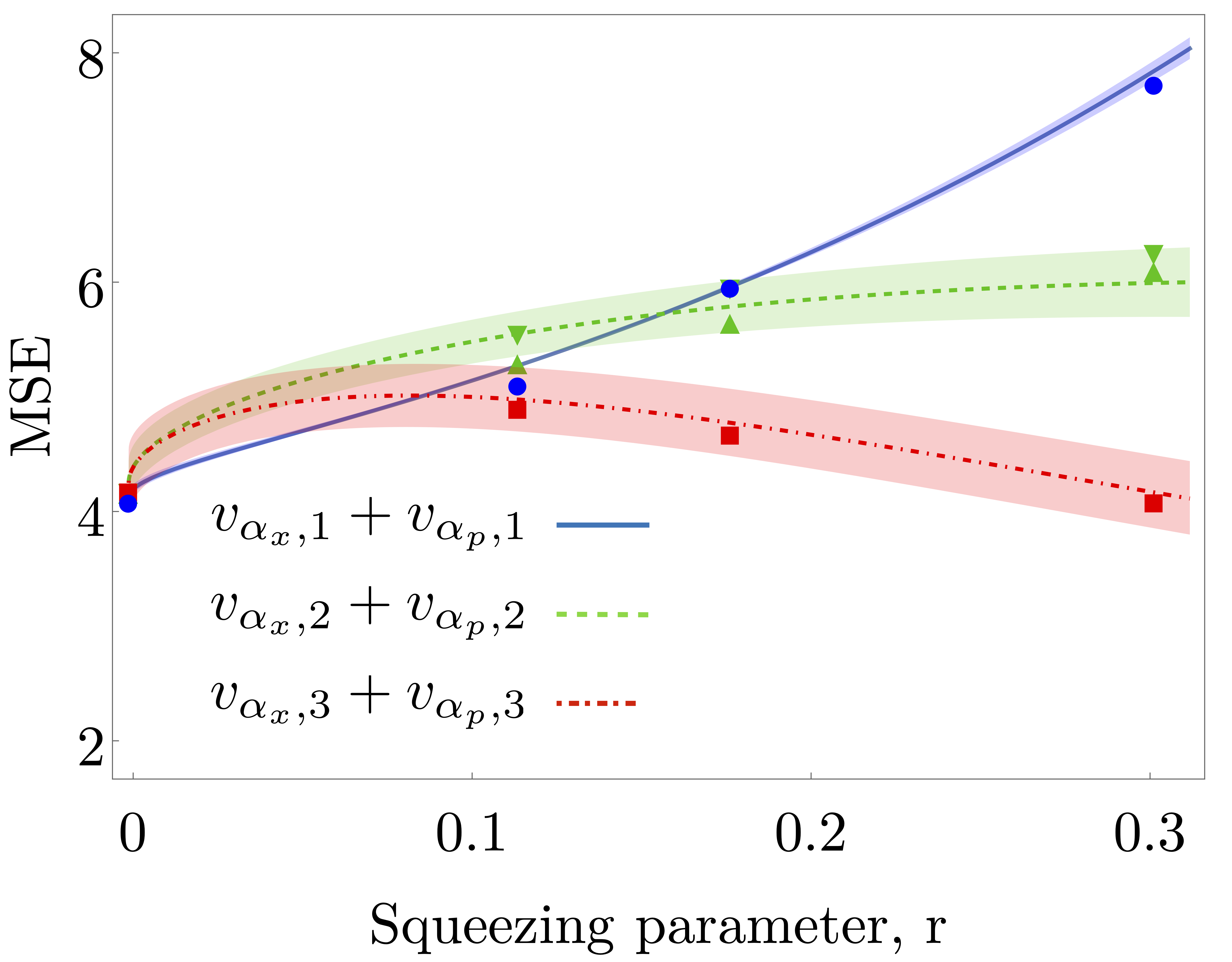}
\caption{\textbf{MSE attainable by different sets of parties working together.} We show the MSE attained experimentally for simultaneously estimating the displacements $(\alpha_x,\alpha_p)$. The solid blue line corresponds to the HCRB, Eq.~\eqref{eqhcrbmain}, which sets the ultimate limit on the MSE which can be attained by any party working individually. The blue data points are the HCRB inferred from the experimentally reconstructed covariance matrix. The dashed green line and data points correspond to the MSE attainable when two parties work together. The upward and downward pointing green triangular data points correspond to parties A and C, and parties A and B working together respectively. The dot-dashed red line and data points correspond to the MSE when all three parties work together. The theoretical lines are based on fitting to a model. The shaded region represents 3\% fluctuations in the optical loss on each arm and thermal noise fluctuations of 0.02 shot noise units on each arm. The displacements being estimated on average have $\abs{\alpha}=0.2$. Each data point corresponds to at least $10^7$ measurement results and statistical error bars based on one standard deviations are smaller than the marker size.}
\label{fig:exp2}
\end{figure}

We next perform an entanglement witnessing task, to ensure that no party is intercepting the quantum state and sending on a different state. The requirement on a CV state to be entangled, see Refs.~\cite{simon2000peres,duan2000inseparability}, allows us to design an entanglement-enabled quantum metrology task. %We shall denote $\hat{x}$ quadrature measurement results for the $i$th party as $x_i$, and similarly for the $\hat{p}$ quadrature.
When all three parties work together, we can construct the following quantities $x_-=x_A/\sqrt{2}-(x_B-x_C)/2$ and $p_+=p_A/\sqrt{2}+(p_B-p_C)/2$, which are unbiased estimators for $(\alpha_x,\alpha_p)/\sqrt{2}$. In the ideal case, using these estimators, it is possible to achieve a MSE for estimating $(\alpha_x,\alpha_p)/\sqrt{2}$ of
\begin{equation}
\label{eq:erent}
v_{\alpha_x'}+v_{\alpha_p'}=4e^{-2r}\;,
\end{equation}
see Appendix~\ref{apenderivest} for the derivation. If the initial two mode state created by the dealer is not entangled, then $v_{\alpha_x'}+v_{\alpha_p'}\geq4$, where we use $v_{\alpha_x'}$ to denote that we are considering the MSE in estimating $\alpha_x/\sqrt{2}$ and similarly for $v_{\alpha_p'}$. We can therefore use this task to verify that we are using a non-classical resource. Our experimental results when using this estimator are shown in Fig.~\ref{fig:exp1}.

\begin{figure*}[t]
\includegraphics[width=\textwidth]{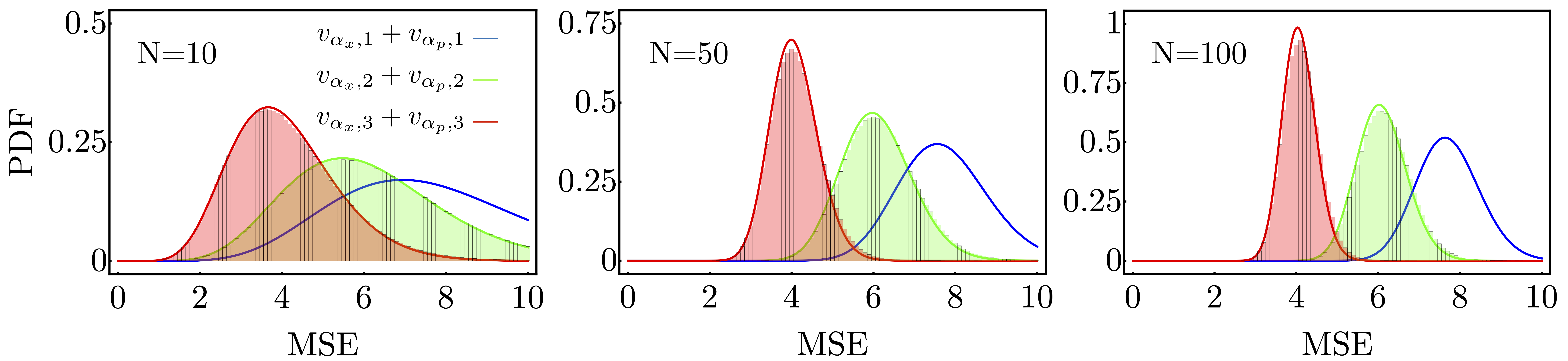}
\caption{\textbf{Probability density functions (PDFs) for the mean squared error (MSE) attainable by different groups of parties using differing numbers of probe states.} The red, green and blue PDFs show the theoretical distribution of MSE values attainable when three parties work together, two parties work together and one party works alone respectively. $N$ is the number of probe states used for estimating both $\alpha_x$ and $\alpha_p$. The histograms show the corresponding experimental data based on $1.85\times10^7/N$ data points. The theoretical PDFs are inferred from Eq.~\eqref{eq:pdf}, using the rightmost experimental data points from Fig.~\ref{fig:exp2}.}
\label{fig:pdfs}
\end{figure*}
\subsubsection{Quantum metrology results}

In Fig.~\ref{fig:exp2} we present the MSE in estimating $(\alpha_x,\alpha_p)$ for different combinations of parties working together. For two and three parties working together, the MSE is obtained directly from the experimental data. The experimental data differs significantly from the ideal theoretical precision (see Fig.~\ref{fig:th1}) due to loss and anti-squeezing present in our experiment. For a single party working alone, the experimental data presented corresponds to the inferred HCRB. In reality, the homodyne detection which we implemented is not sufficient to reach the HCRB. Nevertheless, from the homodyne statistics we can infer the HCRB. We present the inferred HCRB, as opposed to the MSE obtained from homodyne detection, to place limits on the information which could have been extracted by any potentially omnipotent party working alone.
 %Any slight difference between the theoretical fits and data points is likely due to the fact that the data points were taken in different experimental runs, which may correspond to slightly different loss and thermal noise parameters.

\subsubsection{Secure access protocol}
\label{subsecSAP}
We now examine the security of our experiment in the sense of Eqs.~\eqref{eq:security1} and \eqref{eq:security2}.  Fig.~\ref{fig:pdfs} shows the probability density functions (PDFs) for the distribution of MSEs which could be obtained in all three scenarios (one, two and three parties working together) based on the rightmost data points in Fig.~\ref{fig:exp2}. The theoretical PDFs are obtained using Eq.~\eqref{eq:pdf} and the experimentally observed MSEs. The histograms show the experimental data analysed using different numbers of probe states. The slight deviation between the theoretical PDF and the observed PDF is potentially caused by the fact that the experimental MSEs in the $\hat{x}$ and $\hat{p}$ quadratures are not identical, which is assumed when deriving Eq.~\eqref{eq:pdf}. It is evident that, as more probe states are used, the overlap of the distributions of MSE values attainable with multiple parties overlap less with the distribution of MSE values attainable by an individual party. Let us first compare the MSE attainable by all three parties working together to the MSE attainable by a single party. Choosing $v_\text{T}$ as the point where the two distributions are equally likely\footnote{$v_\text{T}=5.5$ when comparing the three-party MSE to the single party MSE and $v_\text{T}=6.8$ when comparing the two-party MSE to the single party MSE}, using 10, 50 and 100 probe states for each quadrature, we can achieve a security given by $\delta=0.18$, $\delta=0.013$ and $\delta=8\times10^{-4}$ respectively. The corresponding success rates are $P_{s,3}=0.87$, $P_{s,3}=0.989$ and $P_{s,3}=0.9993$ respectively. When we compare the two party MSE to that of a single party, using 10, 50 and 100 probe states for each quadrature, we find $\delta=0.39$, $\delta=0.22$ and $\delta=0.13$ and $P_{s,2}=0.68$, $P_{s,2}=0.81$ and $P_{s,2}=0.89$ respectively.

%The area of overlap between the MSE values attained by all three parties and the MSE values attained by any individual party, using 10, 50 and 100 probe states for each quadrature, is equal to 0.32, 0.02 and 0.001 respectively. Comparing two parties working together and any individual party, these numbers become 0.71, 0.40 and 0.24 respectively.
%
%The probability of mistaking one distribution for another is given by the integral of the overlap of the two PDFs. Using 10, 50 and 100 probe states for each quadrature, these probabilities are approximately 0.28, 0.016 and $2\times10^{-4}$ respectively. 
%
%Two parties
%10 probability = 0.1184
%50 probability = 0.14
%100 probability =
%
%Three parties
%10 probability = 0.06375
%50 probability = 
%100 probability =

Let us now consider a scenario where this type of protocol could be useful from a security viewpoint. One could imagine a trust fund bank account which we want to be accessible when either two or three parties work together and inaccessible when parties work individually. By distributing $N$ quantum probe states with unknown displacements $(\alpha_{x,i},\alpha_{p,i})$, a bank manager could only allow access to the bank provided the MSE was below some threshold. Depending on the threshold chosen, either two or three parties would be required to work together to access the bank account. In this manner either a (2,3) or (3,3) access structure can be created with security guaranteed up to the probabilities discussed above.

\subsubsection{Mutual information}

\begin{figure*}[t]
\includegraphics[width=\textwidth]{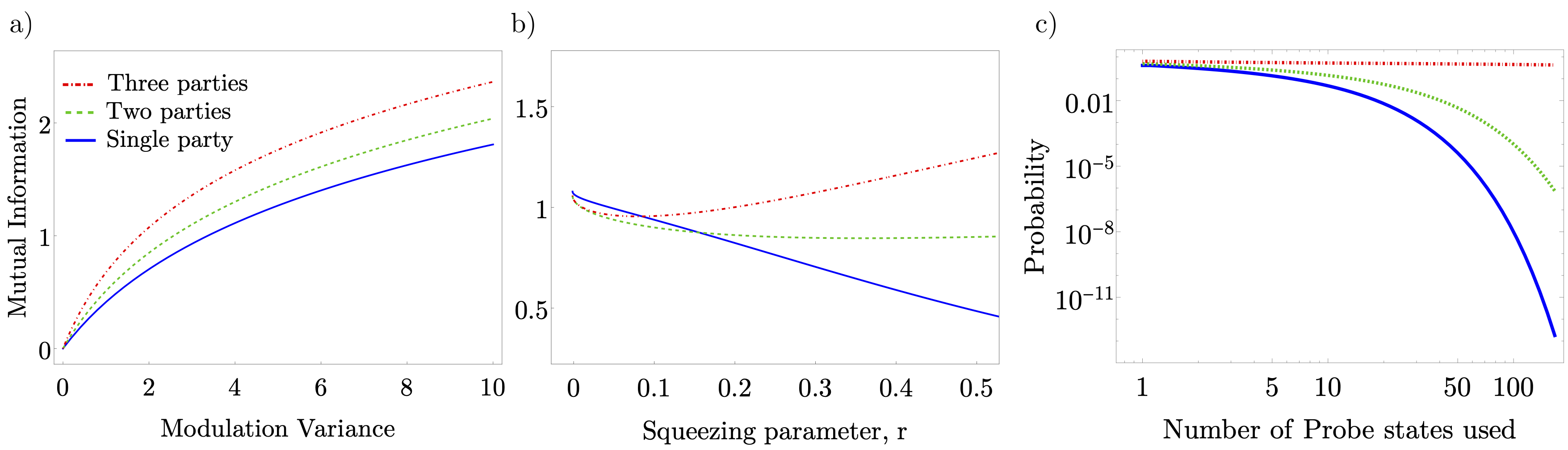}
\caption{\textbf{Mutual information between the dealer and different numbers of parties working together.} a) The dot-dashed red, dashed green and solid blue lines show the mutual information which would be attainable for different input modulation variances when three parties work together, two parties work together and one party works alone respectively. In our experiment, we did not actually implement any random modulations. However, the plot is based on the experimental parameters of the rightmost data points in Fig.~\ref{fig:exp2}. b) The mutual information which can be attained for $V_\text{dist}=2$ as a function of the squeezing. This plot is also based on the experimental parameters. Note that the blue line assumes that any individual party is capable of reaching the HCRB. c) Probability of obtaining a mutual information of more than 1 bit, assuming $V_\text{dist}=4$ based on the rightmost data points in Fig.~\ref{fig:exp2}.}
\label{fig:MI}
\end{figure*}

Finally, we shall discuss how this protocol could be extended to sharing a continuous stream of information. As discussed in section~\ref{sec:attainMI}, we can imagine that the dealer chooses $(\alpha_{x,i},\alpha_{p,i})$ from a Gaussian distribution with variance $V_\text{dist}$. Then the correlation matrix between the dealer and any number of collaborating parties is given by Eq.~\eqref{eqcor}. From this the mutual information between the dealers input and the final estimate of $(\alpha_x,\alpha_p)$ can be calculated using Eq.~\eqref{eq:MI1}. We emphasise that we have not actually implemented this protocol, as in our experiment, we do not change $(\alpha_{x,i},\alpha_{p,i})$ in every run, rather we use a fixed value throughout the experiment. This is a technical limitation of our experiment which can be avoided and does not change any of the main results or conclusions. In Fig.~\ref{fig:MI} a) and b) we show the mutual information which could have been attained in theory, based on the parameters in our experiment, had we drawn $(\alpha_{x,i},\alpha_{p,i})$ from a Gaussian distribution.

Finally we investigate how the probability of attaining a mutual information above a certain value changes as a function of the number of probe states used. This is shown in Fig.~\ref{fig:MI} c), based on the MSE values obtained experimentally. Note that in the experiment, $v_{\alpha_x,m}\neq v_{\alpha_p,m}$ (although this is approximately true), and so we use the average MSE in Fig.~\ref{fig:MI} c).

\subsubsection{Potential security flaws and issues}
Before concluding, we point out some potential security loopholes and other potential issues. To guarantee security in the above protocols, all three parties will need verify that the state they share is entangled, as in section~\ref{ssEntWitness}. This is easily done by using a small subset of the data to verify entanglement. It will also be important to check that the parties are using unbiased estimators of $(\alpha_x,\alpha_p)$, as otherwise it can be possible to violate the HCRB. This can be easily checked by the dealer using a small fraction of the experiments. We also note that party A could claim to have observed high loss on their mode. In this case, in order to ensure the estimates are unbiased, we need to scale the estimate by the inverse of the loss, which increases the MSE attainable. However, party A may not be telling the truth about the loss on their arm. Nevertheless, this issue can be avoided by aborting the protocol if the loss is too high. Additionally, in order to perform homodyne detection at remote stations, it will be necessary for the different parties to share a common phase reference. One approach for this is to distribute a local oscillator to the different parties, a well established technique previously demonstrated over distances greater than 200 km~\cite{zhang2020long}. Alternatively, the parties could use a local local oscillator~\cite{hajomer2024long}. If the excess noise introduced by the use of a local local oscillator is sufficiently low, it would still be possible to demonstrate quantum-enhanced sensitivity.

%If the parties involved in our protocol perform homodyne detection at remote stations using the same local oscillator, this can introduce security loopholes. Recent advances in QKD using a local local oscillator can overcome this issue~\cite{hajomer2024long}.

Finally, it is important to point out that the HCRB sets a limit on the sum of the MSEs in the local estimation setting when parameters are a priori known to within some range. It stands to reason that by removing this information and assuming the parameters to be estimated are unknown no better estimation is possible. The HCRB also in general only applies in the limit of a large number of probe states. This isn't an issue, as we can simply scale all the $(\alpha_{x,i},\alpha_{p,i})$ down by a factor $\sqrt{N}$, and send $N$ copies of this state. Then we can use $N$ large enough for the HCRB to be applicable. In this case, everything from above holds true. A problem with this is it requires $N$ times more channel uses to send the same number of bits of information.

\section{Conclusion}
\label{sec:conc}
We have theoretically and experimentally examined the role of quantum metrology in a secure access protocol and a quantum communication protocol. We envisage these results to be of importance to both the quantum metrology and secure quantum communication communities. In particular, the use of tools from quantum metrology to bound the security of quantum secret sharing may help to connect these two areas and may prompt the search for a more fundamental connection between these two areas. Along this line, Hayashi and Song have recently shown a connection between quantum secret sharing and symmetric private information retrieval~\cite{hayashi2022unified}. Furthermore, as our multiparameter estimation involves multiple distinct parties, it may be of relevance for distributed quantum sensing~\cite{liu2021distributed} or other scenarios where a remote parameter is being probed, such as gravitational field sensing~\cite{conlon2022enhancing}. Our sensing protocol may also be beneficial for the remote sensing of confidential information, such as medical information, as has been discussed for secure quantum enhanced sensing~\cite{huang2019cryptographic,xie2018high,shettell2022cryptographic,takeuchi2019quantum,okane2021quantum,peng2022trusted,shettell2022quantum,moore2023secure,yin2020experimental} (see also Refs.~\cite{giovannetti2002positioning,komar2014quantum}). The fact that our protocol and the corresponding analysis naturally incorporates finite size effects, as discussed in section~\ref{subsecSAP}, is another practical benefit compared to conventional quantum secure communication protocols, where finite size effects have to be added in~\cite{leverrier2010finite,scarani2008quantum}.

There are many ways to extend this research. In the limit of infinite squeezing, our protocol becomes perfectly secure. This suggests that in a DV setting it may be possible to demonstrate perfectly secure secret sharing, with the security guaranteed through quantum metrology. The results in this paper could be strengthened if the dealer implemented displacements on all three modes. Additionally, CV graph states, which have recently been shown to demonstrate an advantage for secret sharing in a quantum network setting~\cite{walk2021sharing}, may further enhance the performance of our protocol. It could also be beneficial to investigate the MSE which can be attained by a malicious party performing commonly considered attacks in QKD. 

%We also note that this protocol is similar in spirit to Ref.~\cite{huang2019cryptographic}, where quantum metrology was used to bound the probability of an eavesdropper being undetected in a DV scenario.

% {\color{red} Something along the lines of using the finite sample paper to bound MSE in the case with a limited number of copies. Might produce a tighter boudn

% Quantum metrology in the finite-sample regime discusses the connection between metrology and state discrimination / hypothesis testing.
% }

\section*{Methods}
\section*{Data Availability}
The data that support the findings of this study are available from the corresponding author upon reasonable request.

\section*{Code Availability}
The code that support the findings of this study is available from the corresponding author upon reasonable request.

\section*{Acknowledgements}
This research was funded by the Australian Research Council Centre of Excellence CE170100012, Laureate Fellowship FL150100019 and the Australian Government Research Training Program Scholarship. 
This research is supported by A*STAR C230917010, Emerging Technology and A*STAR C230917004, Quantum Sensing. 
\section*{Author Contributions}
L.C. and B.S. contributed equally to this work. L.C., B.S., A.W., J.Z. and J.J. contributed to running the experiment and taking the experimental data. B.S. and L.C. modelled the
supporting theory and performed the numerical analysis.
L.C. wrote the manuscript. All authors contributed to
discussions regarding the results in this paper. S.A. and
P.K.L. supervised the project.
\section*{Competing Interests}
The authors declare no competing interests

\appendix

\section{Holevo Cram\'{e}r-Rao bound for simultaneous displacement estimation using an unbalanced thermal state}
\label{apend:HCRB}
We consider the simultaneous estimation of a displacement in both the $\hat{x}$ and $\hat{p}$ quadratures using the Gaussian state with covariance matrix given in Eq.~\eqref{eq:genthermal}. We follow the approach of Ref.~\cite{bradshaw2018ultimate}, where the calculation of the HCRB for estimating Gaussian displacements~\cite{holevo2011probabilistic,holevo1976noncommutative}, was recast as a semi-definite program. We now provide the solutions to both the primal and dual problem, verifying that our solution is correct. Rather than defining many new terms for the sake of a single appendix, we shall use all of the same terminology and definitions as Ref.~\cite{bradshaw2018ultimate}. We shall use the following basis 
\begin{equation}
\begin{bmatrix}
e_1&e_2
\end{bmatrix}=\begin{bmatrix}
0&\frac{1}{\sqrt{v_{2}}}\\
\frac{1}{\sqrt{v_{1}}}&0
\end{bmatrix}\;,
\end{equation}
which satisfies $\alpha(e_j,e_k)=\delta_{j,k}$. This lets us calculate
\begin{equation}
\mathbb{M}=\begin{bmatrix}
\frac{1}{\sqrt{v_{1}}}&0\\
0&\frac{1}{\sqrt{v_{2}}}\;
\end{bmatrix}
\end{equation}
and
\begin{equation}
\mathbb{D}=\begin{bmatrix}
0&\frac{2}{\sqrt{v_{1}v_{2}}}\\
\frac{-2}{\sqrt{v_{1}v_{2}}}&0\;
\end{bmatrix}\;.
\end{equation}
In order to write our solution, we need to define a basis for $2\times2$ real symmetric matrices. We shall use
\begin{equation}
\mathbb{A}_1=\begin{bmatrix}
1&0\\
0&0
\end{bmatrix},\quad\mathbb{A}_2=\begin{bmatrix}
0&0\\
0&1
\end{bmatrix},\quad\mathbb{A}_3=\begin{bmatrix}
0&1\\
1&0
\end{bmatrix}
\end{equation}
and $\mathbb{A}_j=0$ for $j=4,5,6$. We also define $\mathbb{B}_j=0$ for $j=1,2,3$, $\mathbb{B}_4=\mathbb{A}_1$, $\mathbb{B}_5=\mathbb{A}_2$ and $\mathbb{B}_6=\mathbb{A}_3$. Finally, we have $b=[0,0,0,1,1,0]$. This allows us to provide the solution to the primal and dual problems.

\noindent
\textit{Primal problem}\newline\noindent
The primal problem can be written as
\begin{equation}
v_{\alpha_x}+v_{\alpha_p}=\max_{X}\text{tr}\{XC\}\;,
\end{equation}
subject to $\text{tr}\{XB\}=b_j$ where $X$ is a positive Hermitian matrix. We define
\begin{equation}
C=\begin{pmatrix}
0_2&\mathbb{I}_2\\
\mathbb{I}_2&0_2
\end{pmatrix}\bigoplus0_4\bigoplus\mathbb{C}\;,
\end{equation}
where $\mathbb{I}_d$ is the $d\times d$ identity matrix, $0_d$ is the $d\times d$ zero matrix and $\mathbb{C}=(1+\mathrm{i}\mathbb{D}/2)^{-1}$.

The solution to the primal problem is given by
\begin{equation}
X=X_1\bigoplus X_2\;,
\end{equation}
where 
\begin{equation}
X_1=\begin{pmatrix}
1&0&-2-2n_{1}&0\\
0&1&0&-2-2n_{2}\\
-2-2n_{1}&0&4(1+n_{1})^2&0\\
0&-2-2n_{2}&0&4(1+n_{2})^2\\
\end{pmatrix}
\end{equation}
and
\begin{equation}
X_2=\begin{pmatrix}
0&0&0&0  \\
0&0&0&0  \\
0&0&c&\mathrm{i}\sqrt{cd}\\
0&0 &-\mathrm{i}\sqrt{cd}&d\\
\end{pmatrix}\;,
\end{equation}
where $c=4(1+n_{2})^2/v_2$ and $d=4(1+n_{1})^2/v_1$. The non-zero eigenvalues of $X_1$ are 
\begin{equation}
5+4n_{1}(2+n_{1})
\end{equation}
and
\begin{equation}
5+4n_{2}(2+n_{2})\;.
\end{equation}
The only non-zero eigenvalue of $X_2$ is given by $c+d$. Therefore all the eigenvalues of $X$ are non-negative. It is easily verified that the other condition on $X$, $\text{tr}\{XB\}=b_j$, is satisfied by this solution. We can then verify that $\text{tr}\{XC\}=4+2n_{1}+2n_{2}$.

\noindent
\textit{Dual problem}\newline\noindent
By showing that the dual problem has the same solution, we confirm the optimality of our result. The dual problem can be written
\begin{equation}
v_{\alpha_x}+v_{\alpha_p}=\min_yy^Tb\;,
\end{equation}
subject to $\sum_jy_jB_j\geq C$. A solution is given by 
\begin{equation}
y=[\frac{v_{2}}{2+2n_{2}},\frac{v_{1}}{2+2n_{1}},0,2+2n_{1},2+2n_{2},0]\;.
\end{equation}
The matrix $\sum_jy_jB_j-C$ is given by
\begin{equation}
\sum_jy_jB_j-C=Y_1\bigoplus Y_2\bigoplus Y_3\;,
\end{equation}
where
%\begin{widetext}
\begin{equation}
Y_1=\begin{pmatrix}
2(1+n_{1})&0&1&0\\
0&2(1+n_{2})&0&1\\
1&0&\frac{1}{2+2n_{1}}&0\\
0&1&0&\frac{1}{2+2n_{2}}\\
\end{pmatrix}\;,
\end{equation}
%\end{widetext}
\begin{equation}
Y_2=\begin{pmatrix}
1-\frac{1}{2(1+n_{2})}&0\\
0&1-\frac{1}{2(1+n_{1})}
\end{pmatrix}\;,
\end{equation}
and
%\vspace{2cm}
\begin{widetext}
\begin{equation}
Y_3=\begin{pmatrix}
\frac{1}{2(1+n_{2})}+\frac{1}{2n_{1}+2n_{2}+4n_{1}n_{2}}&\frac{-\mathrm{i}\sqrt{(1+2n_{1})(1+2n_{2})}}{2n_{1}+2n_{2}+4n_{1}n_{2}}\\
\frac{\mathrm{i}\sqrt{(1+2n_{1})(1+2n_{2})}}{2n_{1}+2n_{2}+4n_{1}n_{2}}&\frac{1}{2(1+n_{1})}+\frac{1}{2n_{1}+2n_{2}+4n_{1}n_{2}}
\end{pmatrix}\;.
\end{equation}
\end{widetext}
We first consider $Y_1$ and rewrite it as
\begin{equation}
Y_1=\begin{pmatrix}
e&0&1&0\\
0&f&0&1\\
1&0&\frac{1}{e}&0\\
0&1&0&\frac{1}{f}
\end{pmatrix}\;,
\end{equation}
which has eigenvalues
\begin{equation}
\begin{split}
&\frac{1+e^2}{e}\\
\text{and}\qquad&\frac{1+f^2}{f}\;.
%\frac{1}{2}(e+g\pm\sqrt{4+(e-g)^2})\\
%\frac{1}{2}(f+h\pm\sqrt{4+(f-h)^2})\;.
\end{split}
\end{equation}
We can then substitute in from above, to see that there are two non-zero eigenvalues given by
\begin{equation}
\begin{split}
&2(1+n_{1})+\frac{1}{2(1+n_{1})}\\
&2(1+n_{2})+\frac{1}{2(1+n_{2})}\;,
\end{split}
\end{equation}
which are clearly both positive. Similarly it is obvious that both eigenvalues of $Y_2$ are positive. Finally, there is only one non-zero eigenvalue of $Y_3$, given by
\begin{equation}
\frac{1+(2+\frac{n_{2}}{2})n_{2}+n_{1}^2(\frac{1}{2}+n_{2})+n_{1}(2+n_{2}(4+n_{2}))}{(1+n_{1})(1+n_{2})(n_{1}+n_{2}+2n_{1}n_{2})}\;,
\end{equation}
which is guaranteed to be positive. Therefore, the solution $y$ satisfies the constraints, and gives the same solution as the primal problem. Hence, we can be sure our solution is optimal. Therefore, the HCRB for the simultaneous estimation of a displacement in the $\hat{x}$ and $\hat{p}$ quadrature with an unbalanced thermal state is given by
\begin{equation}
\mathcal{C}_\text{H}=4+2n_{1}+2n_{2}\;.
\end{equation}
For equal variances in both quadratures this reduces to the known results of Ref.~\cite{genoni2013optimal,bradshaw2017tight,bakmou2022ultimate}

%\section{Holevo Cram\'{e}r-Rao bound given access to the lost part of both EPR arms}
%Assume now that the dealer makes the TMSV state does the displacements and then sends both arms through a lossy channel to the different parties. The information attained by parties not involved in secret sharing can be bounded by the HCRB. We then get the covariance matrix of this state and use this to bound Eve's information. We can just map to Assad's solution. For the good party we consider the variance attainable by homodyne detection and recombintion

\section{Probability density function for distribution of MSE values}
\label{apen:pdfderiv}
%In our experiments, we use $N$ repetitions for each of the $\hat{x}$ and $\hat{p}$ quadratures. Assuming the MSE in both quadratures is the same, 
We wish to derive the probability density function (PDF) for obtaining a certain MSE given $N$ repetitions of the experiment with mean MSE of $\mu_\text{M}$ in each quadrature. The error in each estimate, $x_i$ or $p_i$, will be randomly distributed following a normal distribution with zero mean and standard deviation of $\sqrt{\mu_\text{M}}$. The quantity we are interested in is the mean of this quantity squared
\begin{equation}
\tilde{\mu}_\text{M}=\frac{1}{N}\sum_{i=1}^{N}x_i^2+\frac{1}{N}\sum_{i=1}^{N}p_i^2\;,
\end{equation}
where $\tilde{\mu}_\text{M}$ is the observed mean MSE. If $\tilde{\mu}_\text{M}$ is repeatedly sampled, the distribution will follow a scaled $\chi^2$ distribution. Assuming that the MSE in both quadratures is the same, we can rewrite $\tilde{\mu}_\text{M}$ as
\begin{equation}
\label{chi2NDOF}
\tilde{\mu}_\text{M}=\frac{\mu_\text{M}}{N}\sum_{i=1}^{2N}x_i^2=\frac{2\mu_\text{M}}{2N}\sum_{i=1}^{2N}x_i^2\;.
\end{equation}
The PDF for a $\chi^2$ distribution with $k$ degrees of freedom is well known and given by
\begin{equation}
P(x)=\frac{1}{2^{k/2}\Gamma(k/2)}(x)^{k/2-1}\text{e}^{-\frac{x}{2}}\;.
\end{equation}
Recognising Eq.~\eqref{chi2NDOF} as a scaled $\chi^2$ distribution with $2N$ degrees of freedom and using the change of variables formula for PDFs, with the function $g(y)=y\mu_\text{M}/N$, we arrive at the PDF for the observed MSE values
\begin{equation}
P(\text{MSE}=x)=\frac{2N}{\mu_\text{M}2^{N}\Gamma(N)}\bigg(\frac{2xN}{\mu_\text{M}}\bigg)^{N-1}\text{e}^{-\frac{xN}{\mu_\text{M}}}\;.
\end{equation}

\section{Derivation of Eq.~\eqref{eq:erent}}
\label{apenderivest}
In this appendix we show that the estimator described in section~\ref{ssEntWitness} for estimating $(\alpha_x,\alpha_p)/\sqrt{2}$, can achieve a MSE given by Eq.~\eqref{eq:erent}. To do this we will use the standard rules for adding random variables. Assume we have three random variables $X$, $Y$ and $Z$ which are drawn from the following multivariate normal distribution
\begin{equation}
\mu=\begin{pmatrix}\mu_X\\ \mu_Y\\ \mu_Z
\end{pmatrix}\qquad\text{and}\qquad
\sigma=\begin{pmatrix}\sigma_X^2&\sigma_{X,Y}&\sigma_{X,Z}\\\sigma_{X,Y}& \sigma_Y^2&\sigma_{Y,Z}\\
\sigma_{X,Z}&\sigma_{Y,Z}& \sigma_Z^2
\end{pmatrix}\;.
\end{equation}
Then it is known that the new variable $aX+bY+cZ$ is normally distributed with mean $a\mu_X+b\mu_Y+c\mu_Z$ and variance $a^2\sigma_X^2+b^2\sigma_Y^2+c^2\sigma_Z^2+2ab\sigma_{X,Y}+2ac\sigma_{X,Z}+2bc\sigma_{Y,Z}$. Using Eq.~\eqref{eqthreemsv} we can calculate the variance of the random variables $x_-=x_A/\sqrt{2}-(x_B-x_C)/2$ and $p_+=p_A/\sqrt{2}+(p_B-p_C)/2$ to be $e^{-2r}$. As we can only assign half of the resources to measuring $x_-$ and half for $p_+$, we scale these variances by a factor of 2. Summing the variance, then shows that the MSE is given by $v_{\alpha_x'}+v_{\alpha_p'}=4e^{-2r}$, in agreement with Eq.~\eqref{eq:erent}. 

Finally, it only remains to show that $x_-$ and $p_+$ are unbiased estimators for $(\alpha_x,\alpha_p)/\sqrt{2}$. As the mean of $x_B$, $x_C$, $p_B$ and $p_C$ are 0, the mean of $x_-$ and $p_+$ are equal to the mean of $x_A/\sqrt{2}$ and $p_A/\sqrt{2}$ respectively. Therefore, the estimator is unbiased.

\bibliography{comp_ref}

\end{document}